\documentclass[12pt]{iopart} \usepackage{graphicx,subfigure}
\usepackage{cite,iopams} \renewcommand{\l}{\left}
\renewcommand{\r}{\right} 
\newcommand{\erfc}{{\rm erfc}}
\newcommand{\avg}[1]{\left\langle{#1}\right\rangle}
\newcommand{\ovl}[1]{\overline{#1}}

\newcommand{\bsy}[1]{\boldsymbol{#1}}
\newcommand{\beq}{\begin{equation}} \newcommand{\eeq}{\end{equation}}
\newcommand{\ii}{{\rm i}}

\begin{document}

\title[]{Von Neumann's expanding model on random graphs}

\author{A De Martino\ddag, C Martelli\P, R Monasson\S,
and I P\'erez Castillo\dag\ddag}

\address{\ddag CNR-INFM (ISC) and Dipartimento di Fisica, Universit\`a
di Roma ``La Sapienza'', p.le Aldo Moro 2, 00185 Roma, Italy}

\address{\P Dipartimento di Scienze Biochimiche, Universit\`a di Roma
``La Sapienza'', p.le Aldo Moro 2, 00185 Roma, Italy}

\address{\S Laboratoire de Physique Th\'eorique de l'ENS, 24 rue
Lhomond - 75231 Paris Cedex 05, France}

\address{\dag Department of Mathematics, King's College London,
Strand, London WC2R 2LS, United Kingdom}

\begin{abstract}
Within the framework of Von Neumann's expanding model, we study the
maximum growth rate $\rho^\star$ achievable by an autocatalytic
reaction network in which reactions involve a finite (fixed or
fluctuating) number $D$ of reagents. $\rho^\star$ is calculated
numerically using a variant of the Minover algorithm, and analytically
via the cavity method for disordered systems. As the ratio between the
number of reactions and that of reagents increases the system passes
from a contracting ($\rho^\star<1$) to an expanding regime
($\rho^\star>1$).  These results extend the scenario derived in the
fully connected model ($D\to \infty$), with the important difference
that, generically, larger growth rates are achievable in the expanding
phase for finite $D$ and in more diluted networks. Moreover, the range
of attainable values of $\rho^\star$ shrinks as the connectivity
increases.
\end{abstract}

\section{Introduction}

Von Neumann's expansion problem was initially formulated to describe
growth in production economies as an autocatalytic process and has
played a key role in the development of the theory of economic growth
\cite{JVN,GaleKT,turnpike}. In a nutshell, it concerns the calculation
of the maximum growth rate achievable by an autocatalytic system of
$M$ reactants (labeled by Greek indices like $\mu,\nu,\dots$)
interconnected by $N$ chemical reactions (labeled by Roman indices
like $i,j,\ldots$) specified by a given stoichiometric matrix. The
basic ingredients are however sufficiently simple to be applicable in
different contexts ranging from economics to systems biology.

In order to state the optimization problem in mathematical terms (see
however \cite{vonno} for a more detailed description), it is
convenient to separate the matrix of input stoichiometric coefficients
$\bsy{A}=\{a_i^\mu\geq 0\}$ (input matrix for brevity) from that of
outputs $\bsy{B}=\{b_i^\mu\geq 0\}$. The relevant microscopic
variables are the reaction fluxes, denoted by $s_i$, which are assumed
to be non-negative. The problem amounts to finding a vector
$\bsy{s}=\{s_i\ge 0\}$ of positive fluxes and a number $\rho> 0$ such
that $\rho$ is maximum subject to
\begin{equation}\label{conda}
\bsy{B}\bsy{s}\ge \rho\bsy{A}\bsy{s}
\end{equation}
where the inequality is to be understood component-wise {\em i.e.}
valid for each reactant $\mu$. The trivial null solution
$\bsy{s}=\bsy{0}$, {\it viz.} $\{s_i=0~\forall i\}$, is not accepted:
at least one component of $\bsy{s}$ must be strictly positive.

Condition (\ref{conda}) requires that the total output of every
reactant is at least $\rho$ times the total input. So the maximum
feasible $\rho$, $\rho^\star$, measures the largest uniform growth
rate achievable by the given set of reactions.  If $\rho^\star<1$, the
optimal state of the system is a contracting one. Otherwise for
$\rho^\star>1$ it is expanding.  The case $\rho^\star=1$ describes
instead a system in which at optimality reaction fluxes are arranged
so as to guarantee mass balance.

Von Neumann's problem has been studied recently in a statistical
mechanics perspective under the assumptions that $M$ scales linearly
with $N$ (with $N\to\infty$) and that the stoichiometric matrices
$\bsy{A}$ and $\bsy{B}$ have quenched random independent and
identically distributed elements \cite{vonno}. With a fully connected
network of reactions (that is, one in which every reaction uses each
reactant both as an input and as an output), one finds a transition
from a contracting to an expanding regime when the parameter $n=N/M$
exceeds the critical value $n_c=1$.

\begin{figure}
\begin{center}
\includegraphics*[width=9cm,angle=0]{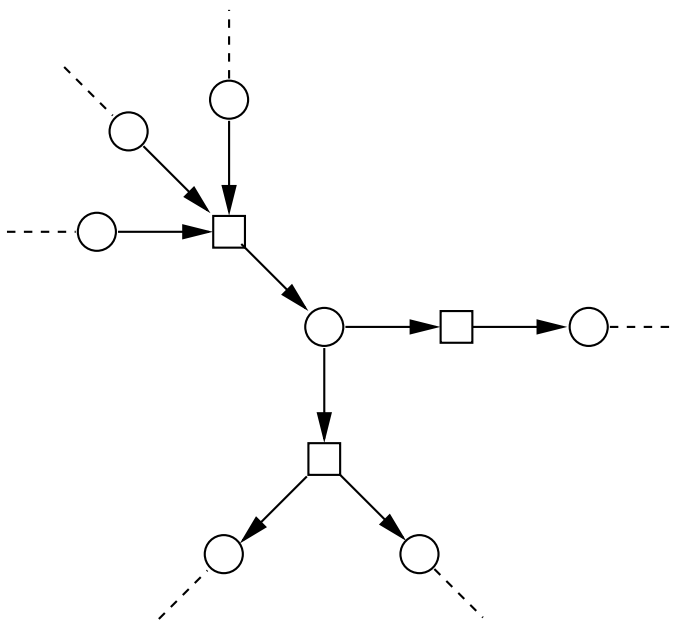}
\end{center}
  \caption{Factor graph representation of a network of reactions and
  reagents. Reactions are represented by circles, reagents by
  squares. Stoichiometric coefficients label the directed
  links. \label{figuno}}
\end{figure}

In this work we extend the analysis of \cite{vonno} to the finitely
connected case where reactions use a finite number of inputs to
produce a finite number of outputs. By analogy with integer
programming problems (see e.g. \cite{ksat}), we represent our
autocatalytic network as a bipartite (factor) graph (see
Fig. \ref{figuno}).
We denote reagents as squares and reactions as circles.  Each circle
has a number of incoming (resp. outgoing) connections to squares,
representing the inputs (resp. outputs) of the reaction, and each of
these links carries the input (resp. output) stoichiometric
coefficient. Similarly, each square has a number of incoming
(resp. outgoing) connections to circles, denoting the reactions where
that reagent enters as an output (resp. input). The notation $i\in\mu$
identifies a reaction $i$ in which a reactant $\mu$ is involved either
as an input or as an output (and vice-versa for $\mu\in i$). To each
reaction node a variable $s_i\geq 0$ is attached, representing the
flux of reaction $i$. Each reactant node $\mu$ carries instead a
function $c^\mu$ given by
\begin{equation} \label{func}
c^\mu=\sum_{i\in\mu}s_i (b_i^\mu-\rho a_i^\mu)
\end{equation}

We are interested in finding non-null flux configurations satisfying
(\ref{conda}), that is such that $c^\mu\geq 0$ for each $\mu$ and
specifically in finding the largest value of $\rho$ for which a
configuration of this type exists. Our approach is developed along two
main lines. On the one hand (Section 2), we employ a suitably modified
Minover algorithm to compute optimal growth rates numerically on any
graph. Results thus derived will be theoretically validated in Section
3, where we use the cavity method to describe solutions analytically
for locally tree-like instances. This approximation turns out to be in
good qualitative agreement with numerical data. The connection of the
latter approach with the fully connected theory developed in
\cite{vonno} is presented in two detailed Appendices.

\section{Numerical analysis}

\subsection{Algorithm}

Formula (\ref{conda}) is reminiscent of pattern storage conditions in
the perceptron, a formal neural network model \cite{neural}. This
analogy allowed us to devise an exact algorithm to solve Von Neumann's
problem on any graph in a time growing only polynomially with the
number $N$ of nodes. The algorithm is an extension of the so-called
Minover algorithm \cite{minover} enforcing the constraint of
positivity for the fluxes, and hereafter referred to as Minover$^+$.

The core part of the algorithm is a subroutine, Minover$^+ (\rho)$, we
now describe. We assume that the matrices $\bsy{A}, \bsy{B}$ are
given, and $\rho$ denotes a real, non-negative number.

\begin{enumerate}
\item Step $\ell =0$: all fluxes are null, $s_i=0$. 
\item Step $\ell +1$: 
Calculate the $M$ functions $c^\mu (\ell)$ defined in (\ref{func})
from the values $s_i(\ell)$ of the fluxes at step $\ell$, and look 
for the reactant with least value,
\begin{equation}
\mu_0(\ell)={\rm arg}~\min_{\mu} c^\mu(\ell) \ ;
\end{equation}
\subitem * if $c^{\mu _0 (\ell)} \geq 0$ then OUTPUT $\bsy{s}(\ell)$
and HALT (the null solution is not accepted); \subitem * otherwise, if
$c^{\mu _0 (\ell)} < 0$, update the fluxes according to the rule
\begin{equation} \label{update}
s_i(\ell+1)=\max\{0,s_i(\ell)+b_i^{\mu_0(\ell)}-\rho\, a_i^{\mu_0(\ell)}\} \ ,
\end{equation}
increase $\ell$ by one, and GO TO (ii).
\end{enumerate}

In case of ties, one may select $\mu_0$ at random uniformly among
those with the same (and lowest) values of $c^\mu$. 

We claim that, if $\rho < \rho ^\star$, the largest growth rate
associated to matrices $\bsy{A}$ and $\bsy{B}$, Minover$^+(\rho)$ will
halt after a finite number of steps, and output a set of non-negative
fluxes guaranteeing a growth rate larger than $\rho$. The proof is
simple and goes as follows.  Assume $\rho < \rho ^\star$. Then there
exists a strictly positive number $\Delta$, hereafter called
stability, and a vector $\bsy{s}^\star\ne\bsy{0}$ of fluxes such
that\footnote{The set ${\cal S}(\rho)$ of fluxes fulfilling
constraints (\ref{conda}) is, by definition of $\rho ^\star$, non
empty. Any vector $\bsy{ s}^\star$ in the interior of ${\cal S}(\rho)$
satisfies (\ref{stab}) for some positive $\Delta$. A natural choice is
the optimal vector of fluxes, {\em i.e.} that associated to $\rho
^\star$. }
\begin{equation} \label{stab}
\bsy{s}^\star\cdot(\bsy{b}^\mu-\rho\bsy{a}^\mu)\geq\Delta\;
|\bsy{s}^\star|~~~~{\rm
for~each~}\mu \ .
\end{equation}
Let 
\begin{equation} \label{defaa}
A = \max _{\mu}  \sum _i  ( b_i^{\mu}-\rho\, 
a_i^{\mu} )^2  > 0\ .
\end{equation}
and consider the functions
$\mathcal{N}(\ell)=\bsy{s}^\star\cdot\bsy{s}(\ell)$ and
$\mathcal{D}(\ell) =| \bsy{s}(\ell) |^2$. We have, from
(\ref{update}),
\begin{eqnarray}
\mathcal{N}(\ell +1 ) &=& \sum _i s_i^\star \;
 \max\{0,s_i(\ell)+b_i^{\mu_0(\ell)}-\rho\, a_i^{\mu_0(\ell)}\}
 \nonumber \\ & \ge & \sum _i s_i^\star \; (s_i(\ell)+
 b_i^{\mu_0(\ell)}-\rho\, a_i^{\mu_0(\ell)})\\& \ge &
 \mathcal{N}(\ell) + \Delta\;|\bsy{s}^\star|\nonumber
\end{eqnarray} 
from the non-negativity of $s_i^\star$ and (\ref{stab}).  We thus
obtain $\mathcal{N}(\ell)\ge \ell \, \Delta\; |\bsy{s}^\star|$ for any
step $\ell$. Turning to function $\mathcal{D}(\ell)$ we get
\begin{eqnarray}
\mathcal{D}(\ell +1 ) &=& \sum _i
\max\{0,s_i(\ell)+b_i^{\mu_0(\ell)}-\rho\, a_i^{\mu_0(\ell)}\}^2
\nonumber \\ & \le & \sum _i ( s_i(\ell)+b_i^{\mu_0(\ell)}-\rho\,
a_i^{\mu_0(\ell)} )^2  \\ &=& \mathcal{D}(\ell) + 2\,
c^{\mu_0(\ell)} + \sum _i ( b_i^{\mu_0(\ell)}-\rho\, a_i^{\mu_0(\ell)}
)^2 \nonumber\\&\le& \mathcal{D}(\ell) + A\nonumber
\end{eqnarray} 
from definition (\ref{defaa}) and the assumption that the algorithm
has not halted at step $\ell$ {\em i.e.} $c^{\mu_0(\ell)} <0$. Hence
$\mathcal{D}(\ell) \le \ell\; A$.  Now let
\begin{equation}
f(\ell)=\frac{\bsy{s}(\ell)\cdot\bsy{s}^\star}{|\bsy{s}(\ell)| \; 
|\bsy{ s}^\star| }
\end{equation}
By  Cauchy-Schwarz inequality $f(\ell ) \leq 1$. But according to the
above calculations
\begin{equation}
f(\ell)=\frac{\mathcal{N}(\ell)}{ |\bsy{s}^\star|
\sqrt{\mathcal{D}(\ell)}} \ge \frac {\Delta}{\sqrt A} \;\sqrt \ell \ .
\end{equation}
Therefore the algorithm halts after 
\begin{equation}\label{lzero}
\ell _0 = \frac{A}{\Delta ^2}
\end{equation}
steps at most. Now call $\ell _s$ the step at which the algorithm
stops.  The halting condition ensures that $c^{\mu_0(\ell _s)} \geq
0$, {\em i.e.}  condition (\ref{conda}) is satisfied, with a non-null
vector of fluxes (by construction the fluxes are non-negative at any
step). Hence $\bsy{s}(\ell _s)$ is a solution to Von Neumann's problem
with growth rate $\rho$.

\begin{figure}
\begin{center}
\includegraphics*[width=8cm]{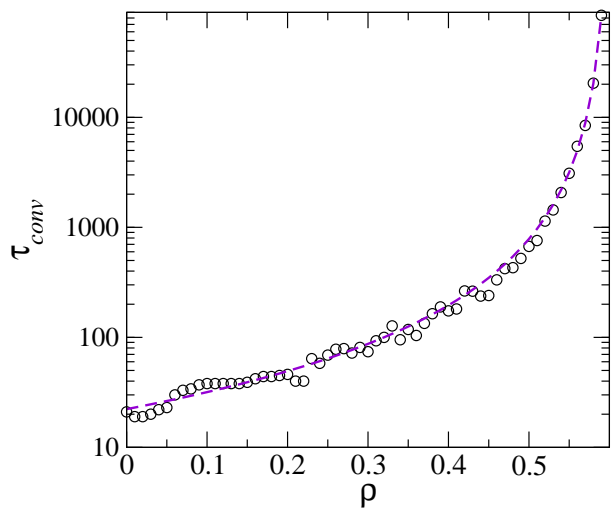}
\end{center}
\caption{Convergence time (in individual steps) versus $\rho$ for a
{\it single} network with $M=100$ reagents and $n=0.66$. The number of
reagents per reaction is fixed at $D=5$, the number of reactions per
reagent is a Poisson random variable with mean $Dn=3.3$. The dashed
line represents the best fit with
$\tau_{conv}\sim(\rho^\star-\rho)^{-2}$. For this network,
$\rho^\star=0.5994$.\label{tempo}}
\end{figure}

Our complete algorithm, Minover$^+$, is an iteration of
Minover$^+(\rho)$ for increasing values of $\rho$. Starting from
$\rho=\epsilon$, a small positive value for which (\ref{conda}) surely
admits a solution \footnote{We can always choose $\epsilon =
\min_{i,\mu} b_i^\mu/a_i^\mu > 0$.}, one can measure the convergence
time (number of steps prior to halting) $\ell_s$ for increasing values
of $\rho$. The convergence time increases too. Now we expect $\Delta $
to vanish as $\rho^\star-\rho$ when $\rho$ gets closer and closer to
its optimal value, thus $\ell_s \sim (\rho^\star-\rho)^{-2}$ from
(\ref{lzero}) (see Fig. \ref{tempo}).
One may then estimate the maximal growth rate by extrapolating from
the log-log plot of the convergence time versus $\rho$.

\subsection{Survey of results}

In what follows we restrict ourselves to purely autocatalytic systems,
that is we assume that every reagent is produced and consumed by at
least one reaction.  Indeed a reagent that only serves as input gives
rise to a constraint of the form
\begin{equation}
c^\mu=-\rho\sum_{i\in\mu}s_i a_i^\mu\geq 0
\end{equation}
which is satisfied by taking $\rho=0$. Such a situation would
immediately force the result $\rho^\star=0$\footnote{The a priori
probability to generate a factor graph without such pathologic
function nodes in the Poissonian model discussed later is given by
$(1-e^{-Dn}(1-e^{-Dn}))^M$, where $M$ is the number of reactants. This
implies that such nodes are always present in the thermodynamic
limit. In this work we discard them completely. However in practical
applications (e.g. metabolic networks \cite{forth}) it is important to
take these nodes into account as prescribed ``sources''
(uptakes).}. On the other hand, ``sink'' reagents (namely those that
are not inputs of any reaction) provide constraints that are
$\rho$-independent and trivially satisfied. Furthermore, for the sake
of simplicity, we fix the number $M$ of reagents and assume that the
quenched random stoichiometric matrix has Gaussian elements with mean
$1$ and variance $1/2$.

\begin{figure}
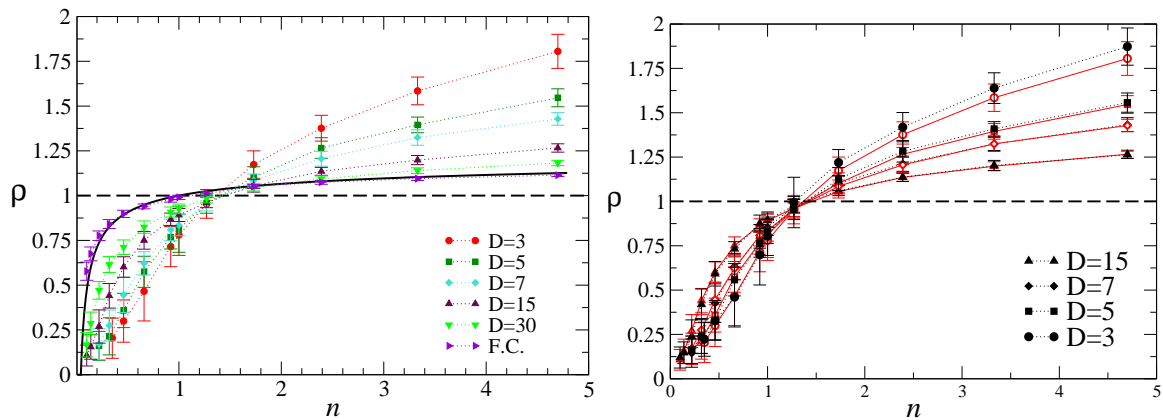

  \begin{center}
  \includegraphics*[width=7.8cm]{cl_rand.eps}
\includegraphics*[width=7.4cm]{cfr_cl.eps}
  \end{center}
  \caption{Left panel: $\rho^\star$ vs $n$ for
  Regular$(N)$/Poisson$(M)$ networks of 100 reagents for various
  $D$. The continuous line is the fully-connected limit for a system
  of the same size \cite{vonno} while F.C. labels numerical results
  for a fully connected system. Right panel: $\rho^\star$ vs $n$ for
  Regular$(N)$/Poisson$(M)$ (open markers) and Regular$(N)$/SF$(M)$
  (closed markers) networks for different values of $D$ (see text for
  details). Averages over 50 samples.\label{rho_regpoi}}
\end{figure}

For a start, we consider Von Neumann's problem on factor graphs in
which every reaction has $D$ inputs and $D$ outputs, while for
reagents we assume that the in(out)-degree distribution is Poissonian,
$P(k)=\lambda^k e^{-\lambda}/k!$ with mean degree $\lambda= Dn$. For
short, we write Regular$(N)$/Poisson$(M)$ to denote this type of
situation. Our analysis focuses on values of $n=N/M>1/D$, which ensure
that the resulting factor graph is connected. Results for $\rho^\star$
in this case are displayed in the left panel of Fig. \ref{rho_regpoi}.
One sees that the overall qualitative behaviour reproduces the results
obtained for the fully connected model. The system passes from an
expanding ($\rho^\star>1$) to a contracting ($\rho^\star<1$) phase
when $n$ is lowered below a critical value. Interestingly, the maximum
growth rate achievable in the diluted system is larger than in the
fully connected model in the expanding phase and higher growth rates
can be achieved in more diluted systems. This conclusion turns out to
be valid generically for all types of networks we studied.

In the right panel of Fig. \ref{rho_regpoi} we compare these results
with those obtained for the case in which the number of reactions per
reagent is power-law distributed (SF or scale-free for short) rather
than Poisson, which enables the coexistence of widely used and rarely
used reagents in the reaction network (reactions still have a
$\delta$-distributed connectivity). More precisely, the number of
reactions per reagent is distributed as $P(k)\sim k^{-\gamma}$ with
$2<\gamma<3$ (as in e.g. metabolic networks \cite{barabba}). The
particular value of the exponent does not affect results in a marked
way. Note that a significant difference (which however lies inside the
error bar) is observed only for the smallest values of $D$.

Next we compare networks with degree regular reactions, where, as said
above, the in/out-degree of reactions is fixed equal to $D$, with
Poisson$(N)$/Poisson$(M)$ networks where also reactions have
fluctuating degree with mean $D$.
\begin{figure}
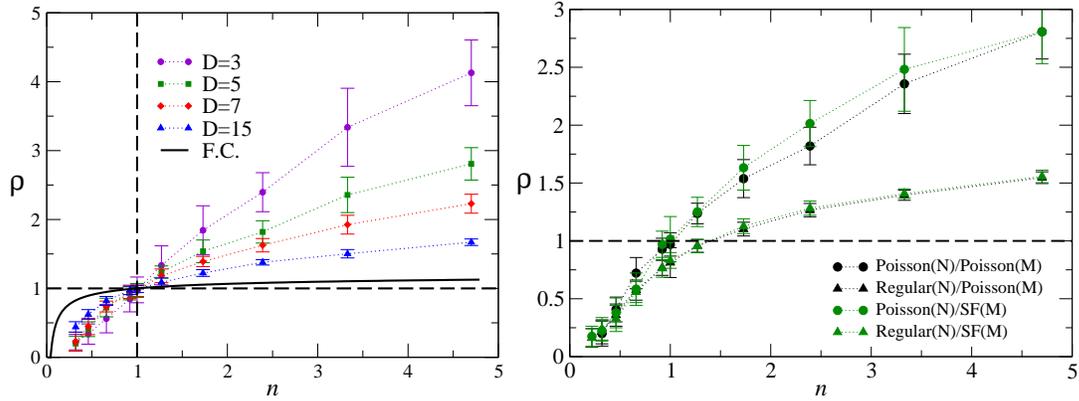

\begin{center}
\includegraphics*[width=6.6cm]{cl-poipoi.eps}
\includegraphics*[width=7.5cm]{cfr_2Poissonx.eps}
\end{center}
\caption{Left panel: $\rho^\star$ vs $n$ for Poisson$(N)$/Poisson$(M)$
networks. Right panel: $\rho^\star$ vs $n$ for different types of
factor graphs with $D=5$. Averages over 50 samples; system of 100
reagents. \label{comp_poisfr}}
\end{figure}
From Fig. \ref{comp_poisfr} one sees that optimal growth rates for all
values of $n$ are always larger in Poissonian networks than in regular
ones, independently of the degree distribution of reagents (right
panel). Moreover a fluctuating connectivity for reactions makes the
expanding regime achievable with fewer reactions (left panel), and in
particular the critical point where $\rho^\star$ becomes larger than 1
is $n_c\simeq 1$, similar to what is found in the fully connected
case. In few words, we can say that topological regularity has a
strong influence on the maximum growth rate achievable and that an
asymmetric choice of degree distributions for reactions and reagents
moves $n_c$ from its fully connected value. In the case above, it
takes a larger repertoire of reactions to sustain an expanding regime.

\begin{figure}
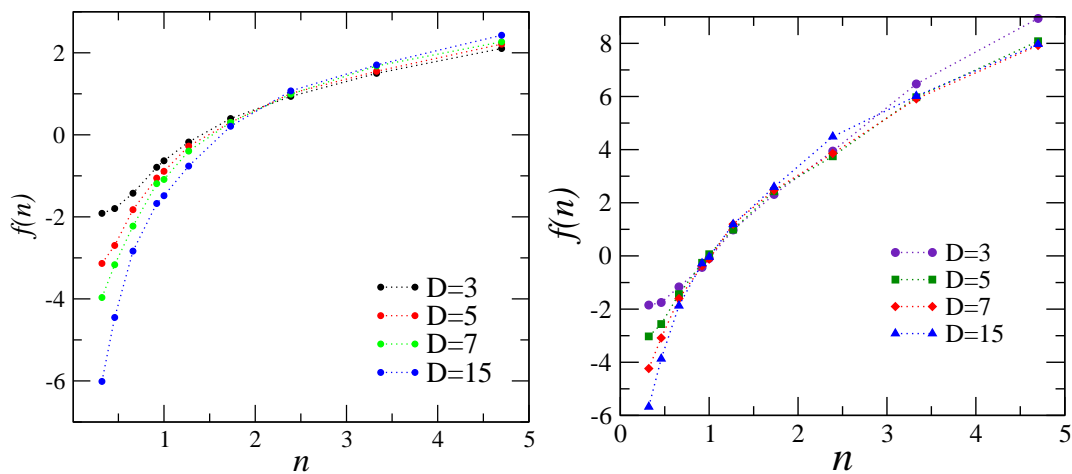

\begin{center}
 \includegraphics*[width=7cm]{scaleRP.eps}
 \includegraphics*[width=7cm]{scalePP.eps}
  \caption{Scaling functions $f(n)$ for Regular$(N)$/Poisson$(M)$
  networks (left) and Poisson$(N)$/Poisson$(M)$ networks
  (right). Error bars not reported.\label{scale}}
\end{center}
\end{figure}

It is possible to find a simple expression that allows to re-scale the
data obtained for different $D$'s onto a single curve (which appears
to be topology-dependent), at least for large $n$. It turns out that
\begin{equation}
\rho^\star(n,D)\simeq \rho^\star_{{\rm F.C.}}(n)+\frac{1}{D}f(n)
\end{equation}
where $\rho^\star_{{\rm F.C.}}(n)$ is the optimal growth rate of the
fully connected system (see Fig. \ref{scale}).
This scaling form, which gives $\rho^\star$'s for diluted systems as
$1/D$ corrections to the fully connected limit, implies that the range
of achievable growth rates shrinks as $D$ increases.

\begin{figure}
\begin{center}
 \includegraphics*[width=13cm]{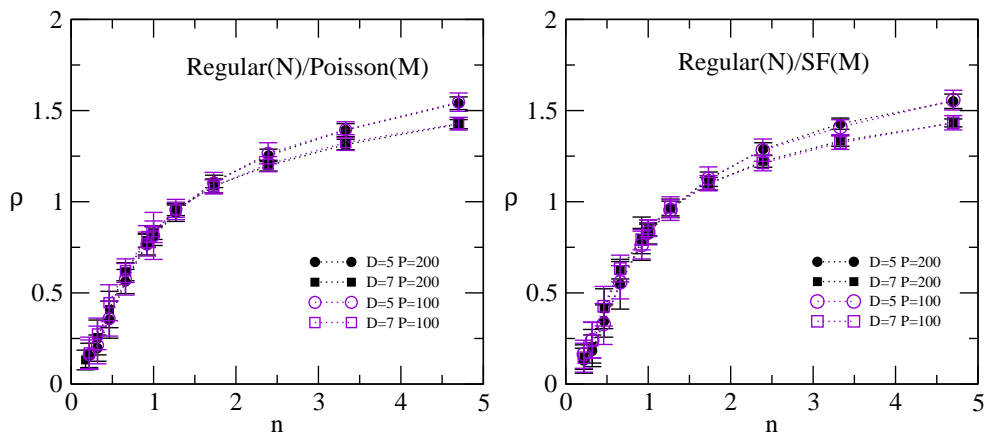}
  \caption{$\rho^\star$ vs $n$ for Regular$(N)$/Poisson$(M)$ and
  Regular($N$)/SF($M$) networks of 100 and 200 reagents for different
  $D$. Averages over 50 samples.\label{size}}
\end{center}
\end{figure}

In Fig. \ref{size}, we show the values of $\rho^\star$ for systems of
different sizes and topologies (similar results hold for other types
of networks).
One sees that already for moderate system sizes finite size effect are
negligible.

\begin{figure}
 \includegraphics*[width=15cm]{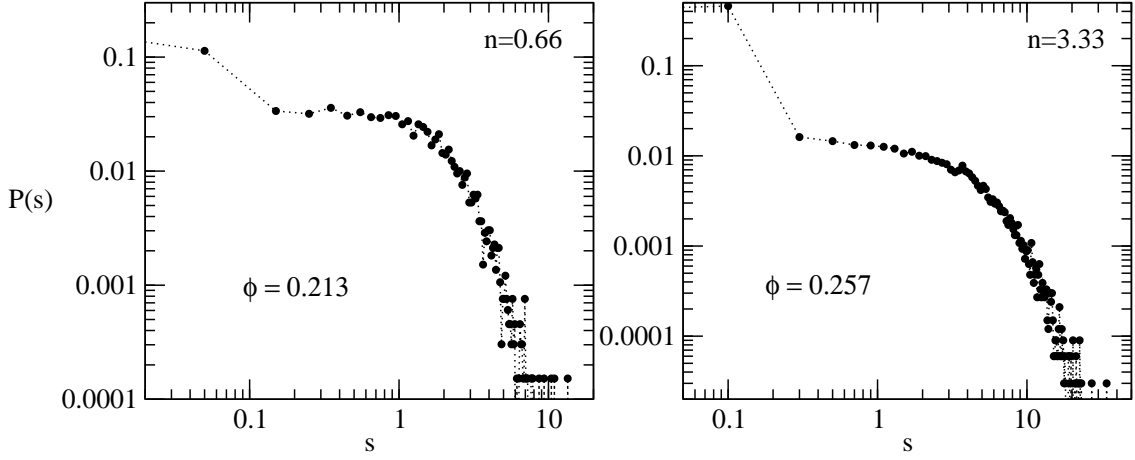}
  \caption{Flux distribution for a Regular$(N)$/Poisson$(M)$ network
  of reagents with $D=5$ inputs and outputs per reaction and different
  values of $n$ (average over 50 samples). System of 200 reagents. The
  $\delta$-peak at $s=0$ (carrying an intensive weight $\phi$ reported
  inside the panels) is not shown.\label{fd1}}
\end{figure}

\begin{figure}
  \includegraphics*[width=15cm]{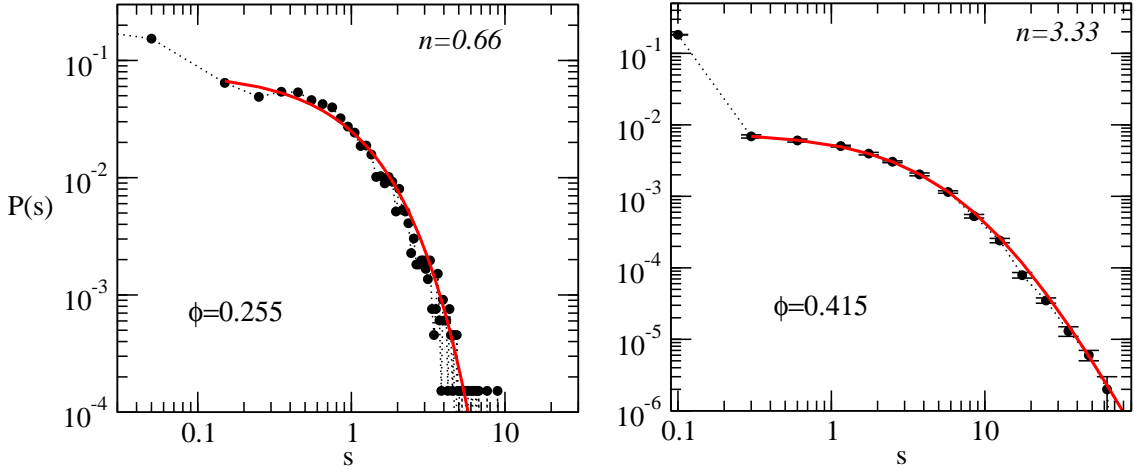}
  \caption{Flux distribution for a Poisson$(N)$/Poisson$(M)$ network
  of reagents with $D=5$ inputs and outputs per reaction and different
  values of $n$ (average over 50 samples). System of 200 reagents. The
  $\delta$-peak at $s=0$ (carrying an intensive weight $\phi$ reported
  inside the panels) is not shown. The red lines represent best fits
  with an exponential distribution of the form $p(x)=A e^{-Bx}$ (left
  panel) and with an algebraic distribution of the form
  $p(x)=A(B+x)^{-C}$ (right panel).  \label{fd2}}
\end{figure}

To conclude, we present (see Figures \ref{fd1} and \ref{fd2}) the flux
distribution at optimality for Regular$(N)$/Poisson$(M)$ and
Poisson$(N)$/Poisson$(M)$ networks for values of $n$ below and above
the critical point.
In the former case, distributions are well fitted by an exponential
form both for small and large $n$. In the latter case, an algebraic
law provides the best fit in the expanding regime. Similar results are
obtained for different connectivity distributions for reaction
nodes. The particular value of the exponent appears to be topology
dependent. However, it is interesting that such distributions arise in
this context, specifically for their relevance in connection to
metabolic networks \cite{forth,barabba}. Note finally that as in the
fully-connected model the weight of the mass at $s=0$ increases with
$n$, though less drastically.

\section{Cavity theory}

Von Neumann's problem possesses the natural cost function
\begin{equation}
E=\sum_{\mu=1}^M\Theta(-c^\mu),
\end{equation}
that, given $\rho$, counts the number of reagents for which the
condition (\ref{conda}) is violated. It is clear that $\rho^\star$
corresponds to the largest $\rho$ for which one can arrange fluxes in
such a way that $E=0$. More generally, we can look for configurations
that minimize $E$, so it is convenient to introduce the ``partition
function''
\begin{equation}
Z=\int \gamma(\bsy{s}) e^{-\beta E} d\bsy{s}\equiv\Tr_{\bsy{s}}e^{-\beta E}
\end{equation}
with $\gamma(\bsy{s})$ some constraint over the configuration space
(e.g. a linear constraint). Minima of the cost function are selected
in the limit $\beta\to\infty$. 

\begin{figure}
\begin{center}
\includegraphics*[width=7cm,angle=0]{factorgraph3.eps}
\quad\quad\quad
\includegraphics*[width=7cm,angle=0]{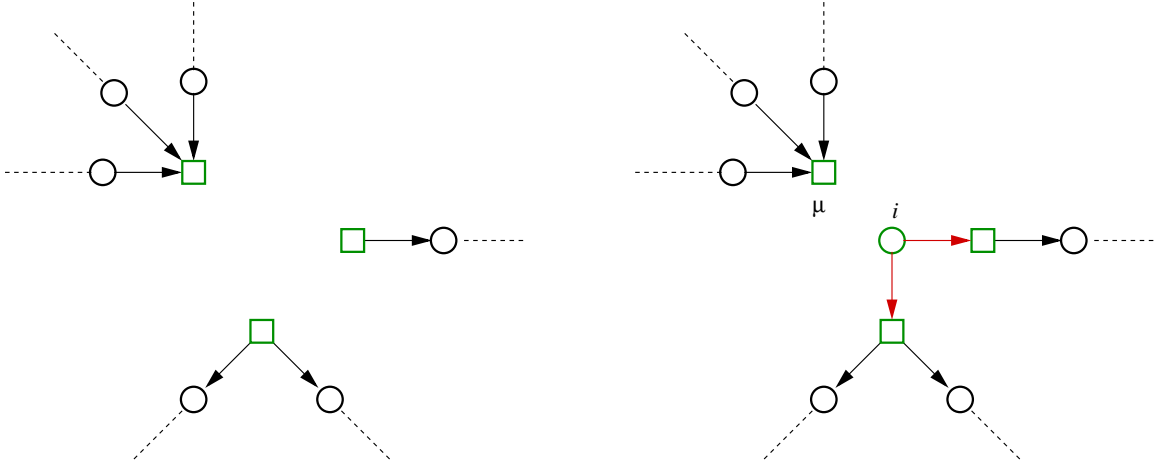}
\end{center}
  \caption{From the graph where reaction $i$ has been removed (left),
  one calculates easily the distribution $p_i^{(\mu)}(s_i)$ in the
  absence of the reagent $\mu$ by merging back all reagents
  participating in reaction $i$ but reagent $\mu$.}
\label{fig:cavity2}
\end{figure}

Let us assume that the factor graph has a locally tree-like structure
\begin{figure}
\begin{center}
\includegraphics*[width=7cm,angle=0]{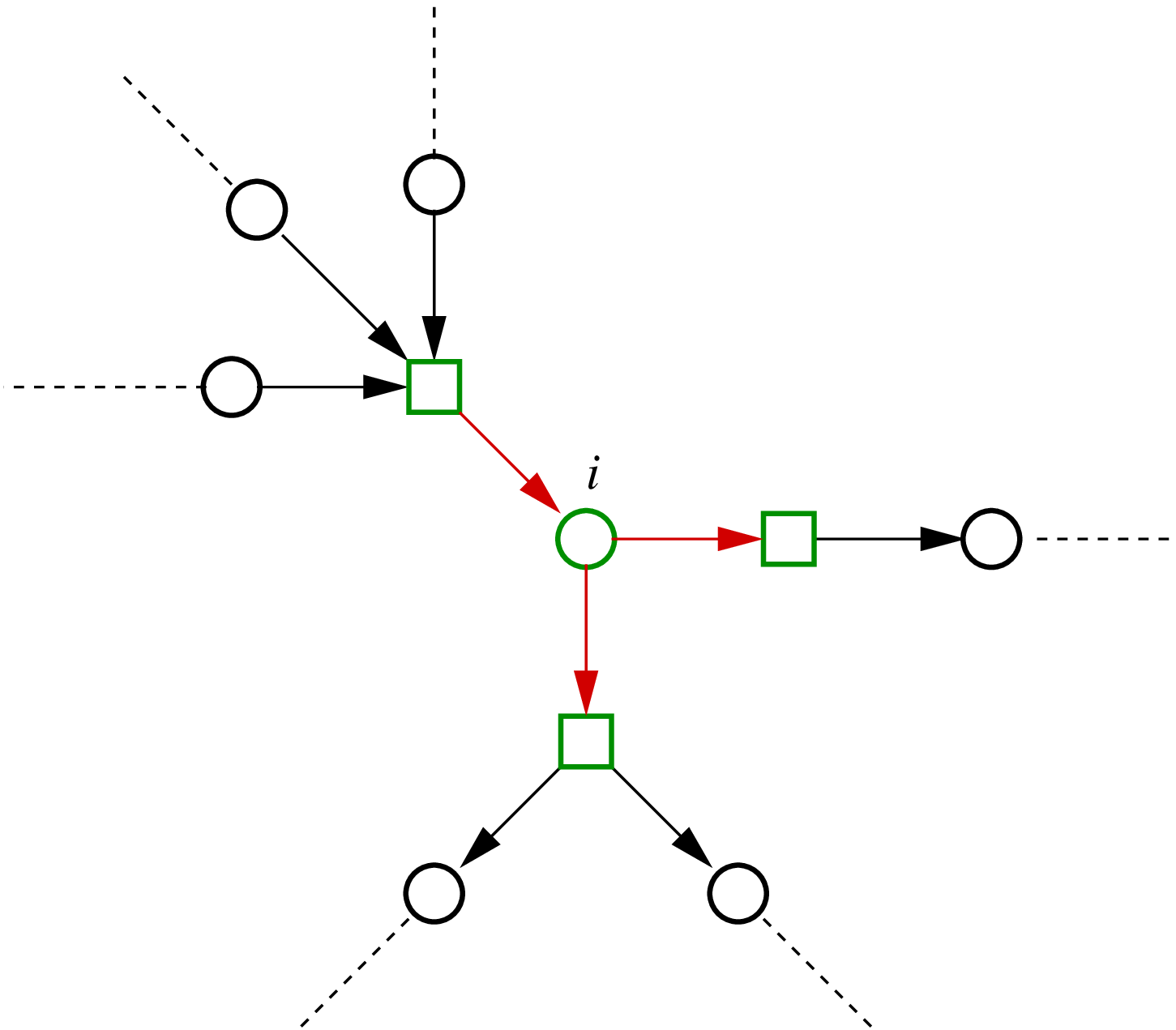}
\quad\quad\quad\includegraphics*[width=7cm,angle=0]{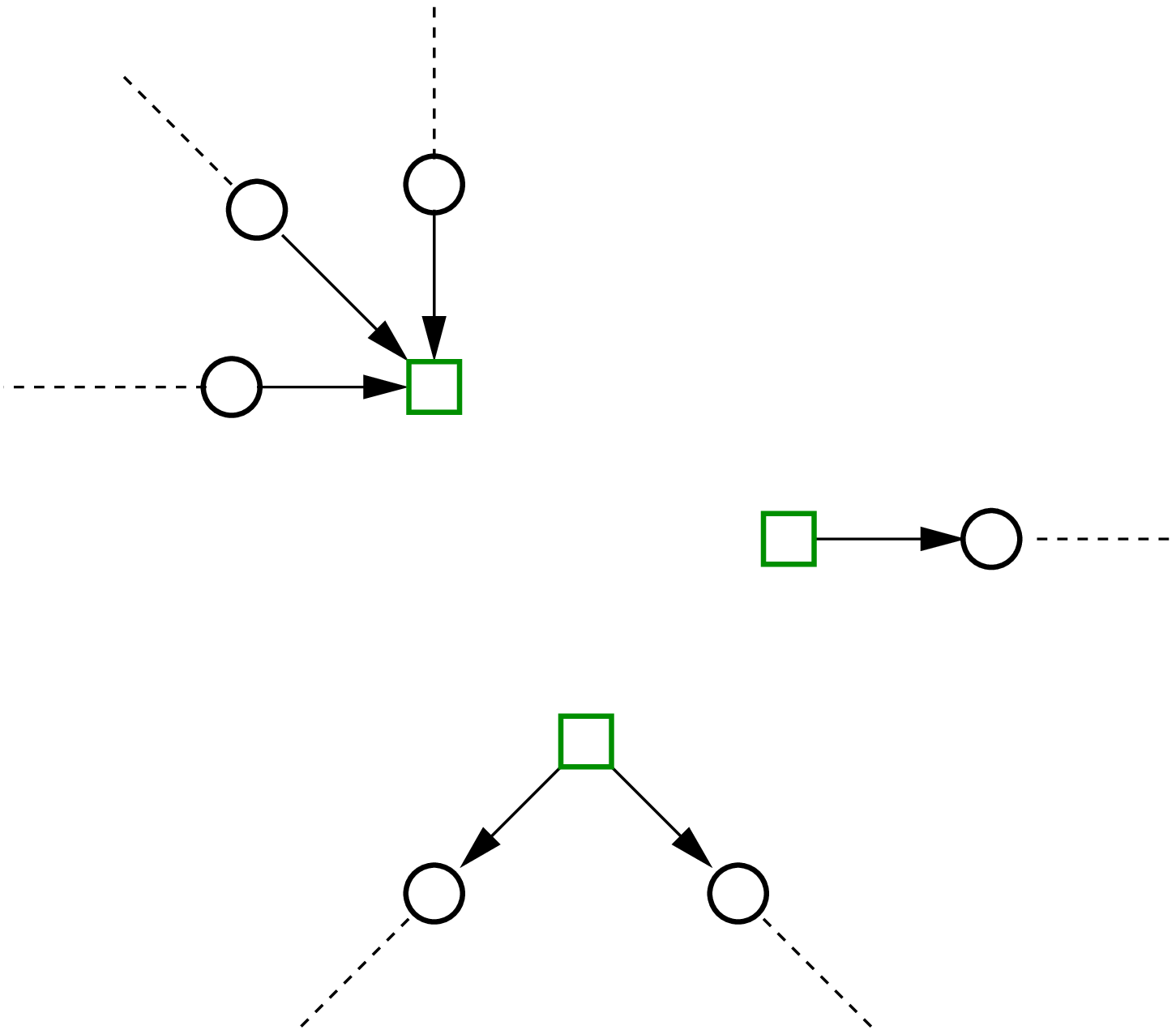}
\end{center}
  \caption{Left: reagents participating in reaction $i$ are
  correlated. Right: If reaction $i$ is removed those reagents become
  uncorrelated on a tree-like graph.}
\label{fig:cavity}
\end{figure}
and consider, as in Figure \ref{fig:cavity}, a reaction $i$ and the
reagents involved in it. It is clear that the joint distribution of
consumptions $\bsy{c}^{\mu\in i}=\{c^\mu\}_{\mu\in i}$ does not
factorize into single-reagent terms, {\it i.e.}
\begin{equation}
q(\bsy{c}^{\mu\in i})\neq\prod_{\mu\in i} q^\mu(c^\mu)
\end{equation}
because the $\mu$'s share the common vertex $i$. However, if we remove
the reaction $i$ (here and in what follows indexes enclosed in
parentheses like $(x)$ denote quantities calculated removing the node
$x$, while sums and products where $x$ is not considered carry the
index $\backslash x$) the resulting joint distribution of consumptions
$q_{(i)}(\bsy{c}^{\mu\in i})$ does factorize:
\begin{equation}
q_{(i)}(\bsy{c}^{\mu\in i})=\prod_{\mu\in i} q_{(i)}^\mu(c^\mu)
\end{equation}
Similarly, with $\bsy{s}_{i\in\mu}=\{s_i\}_{i\in\mu}$, we have
\begin{equation}
p(\bsy{s}_{i\in\mu})\neq \prod_{i\in\mu}p_i(s_i)
\end{equation}
but, removing reagent $\mu$,
\begin{equation}
p^{(\mu)}(\bsy{s}_{i\in\mu})=\prod_{i\in\mu}p_i^{(\mu)}(s_i)
\end{equation}
The quantities $q_{(i)}^\mu(c^\mu)$ and $p_i^{(\mu)}(s_i)$ are known
as {\it cavity distributions}. Now for all $i=1,\ldots,N$ and $\mu\in
i$ it is possible to find equations for the cavity distributions by
simply merging back all disconnected nodes but one (see
Fig. \ref{fig:cavity2}). It is easily seen in particular that
\begin{eqnarray}
p_i^{(\mu)}(s_i)&=&\frac{1}{Z_i^{(\mu)}} \int e^{-\beta\sum_{\nu\in
i\backslash \mu}\Theta\l[-c_{(i)}^\nu-s_i(b_i^\nu-\rho
a_i^\nu)\r]} \prod_{\nu\in i\backslash\mu}q_{(i)}^\nu(c^\nu)
dc^\nu\nonumber\\\label{zoppola}
&\equiv&\frac{1}{Z_i^{(\mu)}}\Tr_{\bsy{c}^{\nu\in i\backslash \mu}}
\l[\prod_{\nu\in i\backslash\mu}q_{(i)}^\nu(c^\nu)\r]
e^{-\beta\sum_{\nu\in
i\backslash \mu}\Theta\l[-c_{(i)}^\nu-s_i(b_i^\nu-\rho
a_i^\nu)\r]}
\end{eqnarray}
(where $Z_i^{(\mu)}$ is a normalization factor ensuring that $\int
p_i^{(\mu)}(s_i)ds_i=1$) while for each $\mu=1,\ldots,M$
and $i\in\mu$ we have
\begin{equation}\label{zoppolo}
q_{(i)}^\mu(c^\mu)=
\Tr_{\bsy{s}_{j\in\mu\backslash i}}
\l[\prod_{j\in\mu\backslash i} p_j^{(\mu)}(s_j)\r]
\delta\l[c^\mu-\sum_{j\in\mu\backslash i} s_j
(b_j^\mu-\rho a_j^\mu)\r]
\end{equation}
Note that these in turn imply that
\begin{eqnarray}
p_i(s_i)=\frac{1}{Z_i}\Tr_{\bsy{c}^{\nu\in i}} 
q_{(i)}(\bsy{c}^{\nu\in i})
e^{-\beta\sum_{\nu\in
i}\Theta\l[-c_{(i)}^\nu-s_i(b_i^\nu-\rho
a_i^\nu)\r]}\\
q^\mu(c^\mu)=
\Tr_{\bsy{s}_{j\in\mu}}
p^{(\mu)}(\bsy{s}_{j\in\mu})
\delta\l[c^\mu-\sum_{j\in\mu}s_j
(b_j^\mu-\rho a_j^\mu)\r]
\end{eqnarray}
so that the whole probability distribution can be reconstructed from
the set of cavity distributions. The statistical average cost function
is finally given by
\begin{eqnarray}
\avg{\frac{E}{N}}&=&\frac{1}{N }\sum_{\mu=1}^M{\rm Tr}_{c} \Theta(-c) 
\frac{q^{\mu}(c)\left[e^{-\beta}+(1-e^{-\beta})
\Theta\left(c\right)\right]}{{\rm Tr}_{c} 
q^\mu(c)\left[e^{-\beta}+(1-e^{-\beta})\Theta\left(c\right)\right] }
\end{eqnarray}
The cavity theory developed here recovers the replica theory of
\cite{vonno} in the fully-connected limit. This is shown in detail in
Appendices A, where a slightly revised replica theory is discussed,
and B, where the fully connected limit is constructed explicitly.

Because the configurational variables $s_i$ and $c^\mu$ are
continuous, solving equations (\ref{zoppola}) and (\ref{zoppolo}),
which in the zero temperature ($\beta\to\infty$) limit take the form
\begin{eqnarray}
p_i^{(\mu)}(s_i)&=&\frac{1}{Z_i^{(\mu)}}\Tr_{\bsy{c}^{\nu\in
i\backslash \mu}} \l[\prod_{\nu\in
i\backslash\mu}q_{(i)}^\nu(c^\nu)\r] \prod_{\nu\in i\backslash
\mu}\Theta\l[c_{(i)}^\nu+s_i(b_i^\nu-\rho
a_i^\nu)\r]\label{eq:cavity1}\\ q_{(i)}^\mu(c^\mu)&=&
\Tr_{\bsy{s}_{j\in\mu\backslash i}} \l[\prod_{j\in\mu\backslash i}
p_j^{(\mu)}(s_j)\r] \delta\l[c^\mu-\sum_{j\in\mu\backslash i} s_j
(b_j^\mu-\rho a_j^\mu)\r]\label{eq:cavity2},
\end{eqnarray} 
by the standard method of population dynamics requires studying a
population of populations, as {\it e.g.} in \cite{ska} (in other
words, it is equivalent to a one-step RSB calculation already at the
replica-symmetric level). We therefore decided to focus our analysis
on the calculation of the optimal growth rate, for which it is
possible to obtain good results at a modest computational cost.

We are thus interested in finding the critical line
$\rho^\star(n)$. One may proceed as follows. Assume there is no linear
constraint (the value of $\rho^\star$ is unaffected by the presence of
a linear constraint). Then the equations always admit the trivial
solution ($\bsy{s}=\bsy{0}$) for all values of $\rho$. In fact, for
$\rho>\rho^\star$ this is the only zero-energy solution while below
$\rho^\star$ other non-null solutions exist.  Thus, starting from
random initial conditions for the fluxes, we would expect the average
flux to vanish (resp. remain different from zero) under the iteration
of the cavity equations for values of $\rho$ above (resp. below)
$\rho^\star$. This obviously assumes that the iteration of the cavity
equations for $\rho<\rho^\star$ keeps the fluxes away from their
trivial value, which turns out to be the case.

\begin{figure}[t]
  \begin{center}
  \includegraphics*[width=9cm]{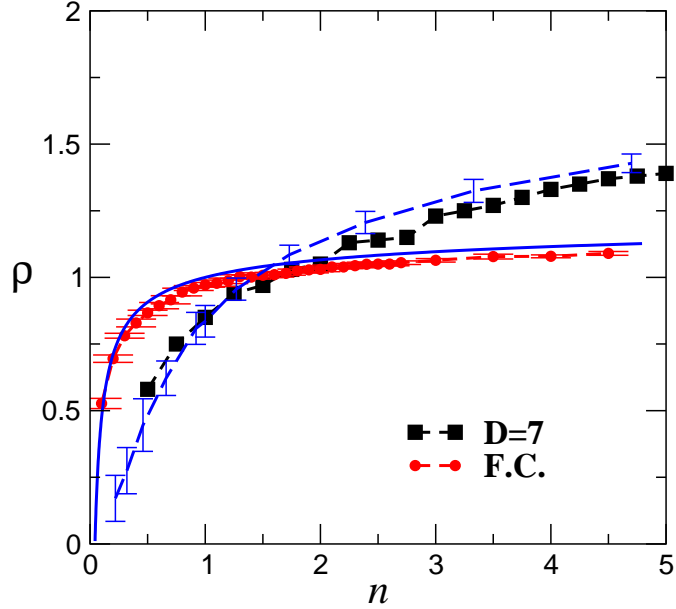}
  \end{center}
  \caption{Results from cavity theory. The continuous (blue) line is
  the theoretical (replica) prediction in the fully connected
  limit. The dashed line represents Minover results for
  Regular($N$)/Poisson ($M$) graphs, with error bars. Markers
  correspond to iteration of the fully connected (circles, average
  over 10 samples, number of reagents $M=100$) and
  Regular($N$)/Poisson ($M$) cavity equations (squares).}
\label{fig: popdyn}
\end{figure}

To check the validity of this assumption, we have first considered the
cavity equations (\ref{eq:cavity1}) and (\ref{eq:cavity2}) in their
fully connected limit (see Appendices for details), \textit{viz.}
\begin{eqnarray}
x_{(i)}^\mu&=&M_{0}\left(h^\mu_{(i)},\phi^\mu_{(i)}\right)\\
 y_{(i)}^\mu&=&M_{1}\left(h^\mu_{(i)},\phi^\mu_{(i)}\right)\\
 m_\ell^{(\mu)}&=&f_1\left( A^{(\mu)}_\ell,B^{(\mu)}_\ell\right)\\
 L_\ell^{(\mu)}&=&f_2\left( A^{(\mu)}_\ell,B^{(\mu)}_\ell\right)
\end{eqnarray}
with
\begin{eqnarray}
h^\mu_{(i)}&=&\frac{1}{\sqrt{K}}\sum_{\ell\in \mu\backslash
i}\xi_\ell^\mu(\rho)m_\ell^{(\mu)}\\
\phi^\mu_{(i)}&=&\frac{1}{K}\sum_{\ell\in \mu\backslash
i}[\xi_\ell^\mu(\rho)]^2[L_\ell^{(\mu)}-[m_\ell^{(\mu)}]^2]\\
A^{(\mu)}_i&=&\frac{1}{\sqrt{K}}\,\sum_{\nu\in i\backslash \mu}
\xi_i^\nu(\rho)x_{(i)}^\mu\\ B^{(\mu)}_i&=&\frac{1}{2K}\sum_{\nu\in
i\backslash \mu} [\xi_i^\nu(\rho)]^2\left(
y_{(i)}^\nu+[x_{(i)}^\nu]^2\right)\\
\xi_i^\mu(\rho)&=&b_i^\mu-\rho a_i^\mu
\end{eqnarray}
and where $M_{s}(a,b)$ and $f_k(a,b)$ are defined as follows
($H(x)=\erfc(x/\sqrt{2})/2$)
\begin{eqnarray}
M_{s}(a,b)&=&\frac{\left(1-e^{-\beta}\right)\left(\frac{\partial}{\partial
c}\right)^{s}\mathcal{G}_c(a,b)\Big|_{s=0}}{e^{-\beta}+
(1-e^{-\beta})H\left(\frac{-a}{\sqrt{b}}\right)}~~,~~~~~(\beta\to\infty)\\
f_k(a,b)&=&\frac{{\rm Tr}_{s} e^{a s-b s^2} s^{k}}{{\rm Tr}_{s} e^{ a
s-b s^2}}
\end{eqnarray}
with $\mathcal{G}_c(a,b)$ a Gaussian distribution for the variable $c$
with mean $a$ and variance $b$. For computational reasons, it is more
convenient to iterate the corresponding TAP equations \cite{shamir},
namely
\begin{eqnarray}
x^\mu&=& M_{0}\left(\frac{1}{\sqrt{K}}\sum_{\ell\in
 \mu}\xi_\ell^\mu(\rho)m_\ell-x^\mu \phi^\mu,\phi^\mu\right)\\
 y^\mu&=& M_{1}\left(\frac{1}{\sqrt{K}}\sum_{\ell\in
 \mu}\xi_\ell^\mu(\rho)m_\ell-x^\mu \phi^\mu,\phi^\mu\right)\\
 m_{i}&=&f_{1}\left(\frac{1}{\sqrt{K}}\,\sum_{\nu\in i}
 \xi_i^\nu(\rho) x^\mu+m_i B_i,B_i\right)\\
 L_{i}&=&f_{2}\left(\frac{1}{\sqrt{K}}\,\sum_{\nu\in i}
 \xi_i^\nu(\rho) x^\mu+m_iB_i,B_i\right)\\
 B_i&=&\frac{1}{2K}\sum_{\nu\in i} [\xi_i^\nu(\rho)]^2\left(
 y^\nu+(x^\nu)^2\right)\\ \phi^\mu&=& \frac{1}{K}\sum_{i\in
 \mu}[\xi_i^\mu(\rho)]^2\left(L_i-m_i^2\right)
\end{eqnarray}
Given a matrix $\xi_i^\mu(\rho)$, we solve the preceding set of
coupled equations by iteration: starting from a small value of $\rho$
we monitor the average flux and determine $\rho^\star$ when, under
iteration of the equations, the average flux goes to zero. In
Fig. \ref{fig: popdyn} we compare the results for a system of $M=100$
reagents (average over 10 samples) with the theoretical
prediction. The agreement is fairly good.

We have then considered the same approach to calculate $\rho^\star$
for a Regular$(N)$/Poisson$(M)$ network with connectivity $D=7$ (see
Fig. \ref{fig: popdyn}). To do so, we have considered the ensemble
version of equations (\ref{eq:cavity1}) and (\ref{eq:cavity2}), which
take the form (we suppress cavity indexes $(x)$ for simplicity)
\begin{eqnarray}
p_i(s)&=&\frac{1}{Z_i}\prod_{\nu=1}^{2D-1}\Bigg[\Tr_{{c}^{\nu}}
q^\nu(c^\nu) \Theta\l[c^\nu+s(b_i^\nu-\rho
a_i^\nu)\r]\Big]\label{eq:cavity1v3}\\ q^\mu(c)&=&
\l[\prod_{j=1}^{k_\mu-1}\Tr_{{s}_{j}} p_j(s_j)\r]
\delta\l[c-\sum_{j=1}^{k_\mu-1}s_j (b_j^\mu-\rho
a_j^\mu)\r]\label{eq:cavity2v3}
\end{eqnarray}
with $k_\mu$ a random Poisson number with average $n D$. In the
ensemble, which is obtained after the limit $N\to\infty$ has been
performed and where $n$ plays the role of an external parameter, the
indexes $i$ and $\mu$ now take values $i,\mu=1,2,3,\ldots$ and
describe an infinite population of probabilities
$\{q^\mu(c)\}_{\mu=1,2,\ldots}$ and $\{p_i(s)\}_{i=1,2,\ldots}$. We
have studied a finite population of probabilities, \textit{i.e.}
$i,\mu=1,\ldots,\mathcal{N}_{pop}$, which are initialised randomly.
In order to find $\rho^\star$, we start from a small initial value for
$\rho$. The population is iterated via equations (\ref{eq:cavity1v3})
and (\ref{eq:cavity2v3}).  If after a certain (sufficiently large)
number of iteration steps the average flux is different from zero, we
increase the value of $\rho$ by a small value and restart the
procedure. Eventually we can locate a value of $\rho$ at which the
average flux vanishes. This defines $\rho^\star$. In Fig. \ref{fig:
popdyn} we display the resulting curve $\rho^\star(n)$ for a
population of $\mathcal{N}_{pop}=50$ distributions and $D=7$, which
qualitatively agrees with the results found by the Minover$^+$
algorithm. (We remind that cavity equations are obtained for a
tree-like topology and the network we considered in the numerical
solution is relatively small.)

\section{Summary}

Despite evident quantitative differences, the overall picture obtained
for the fully-connected Von Neumann's model is roughly preserved in
the case of finite connectivity, discussed here. The major difference
appears to be that much larger growth rates can be achieved in diluted
networks (reactions with few inputs/outputs are evidently more
efficient) and, generically, when reactions have stochastic
connectivities. The framework of Von Neumann's problem, including its
global optimization aspect, is sufficiently simple to ensure that the
methods developed in this work can be applied in several different
contexts, particularly in economics and biology. Work along both lines
is in progress.

\ack

It is our pleasure to thank G Bianconi, S Cocco, E Marinari, M
Marsili, A Pagnani, G Parisi, F Ricci Tersenghi and T Rizzo for
valuable comments and suggestions.

\appendix

\section{The fully-connected model revisited}

We shall briefly re-consider here the replica approach of \cite{vonno}
for the fully connected model in the Derrida-Gardner framework (as
opposed to the Gardner one employed in \cite{vonno}). This allows us
to derive equations for the order parameters that are easily compared
with those which shall be derived in the following section as the
fully connected limit of the cavity theory.

Our starting point is the partition function, which we write in this
case
\begin{equation}\fl
Z=\Tr_{\bsy{c}}\prod_{\mu=1}^M \left[e^{-\beta}+(1-e^{-\beta})\Theta
(c^\mu)\right] \Tr_{\bsy{s}} \prod_{\mu=1}^M
\delta\left(c^\mu-\frac{1}{\sqrt{N}}\sum_{i=1}^N
s_i\xi_{i}^\mu(\rho)\right)
\end{equation}
where
\begin{equation}\label{defa}
\xi_i^\mu(\rho)=b_i^\mu-\rho a_i^\mu
\end{equation}
As in \cite{vonno}, we write $b_i^\mu=\ovl{b}(1+\beta_i^\mu)$ and
$a_i^\mu=\ovl{a}(1+\alpha_i^\mu)$ with Gaussian $\alpha_i^\mu$ and
$\beta_i^\mu$ and further set $\ovl{a}=\ovl{b}=1$. Moreover, we
re-cast $\rho$ as $\rho=1+g/\sqrt{N}$ so that
\begin{equation}\label{defa2}
\xi_i^\mu(\rho)=\beta_i^\mu-\alpha_i^\mu-\frac{g}{\sqrt{N}}(1+\alpha_i^\mu)
\end{equation}
The replicated and disorder-averaged partition function reads
\begin{eqnarray}
\ovl{Z^r}=\Tr_{\{\bsy{c}_1,\ldots,\bsy{c}_r\}}\prod_{\ell=1}^r
\prod_{\mu=1}^M \left[e^{-\beta}+(1-e^{-\beta})\Theta
(c^\mu_\ell)\right]\nonumber\\
\hspace{1cm}\times \Tr_{\{\bsy{s}_1,\ldots,\bsy{s}_r\}}
\ovl{\prod_{\ell=1}^r \prod_{\mu=1}^M
\delta\left(c_\ell^\mu-\frac{1}{\sqrt{N}}\sum_{i=1}^N
s_{i\ell}\xi_{i}^\mu(\rho)\right)}
\end{eqnarray}
If the $\delta$-distributions are written in their Fourier
representation (with multipliers $\widehat{c}_\ell^\mu$) the disorder
average can be carried out easily, generating the standard order
parameters
\begin{equation}
q_{\ell\ell'}=\frac{1}{N}\sum_i s_{i\ell} s_{i\ell'}~~~~~~~
m_\ell=\frac{1}{N}\sum_i s_{i\ell}
\end{equation}
which can be forced into the partition function with proper Lagrange
multipliers $\widehat{q}_{\ell\ell'}$ and $\widehat{m}_\ell$.  One
finally finds
\begin{equation}\label{derp}
\overline{Z^r}=\int D\bsy{q}D\widehat{\bsy{q}}
D\bsy{m}D\widehat{\bsy{m}}~
e^{N[\sum_{\ell\leq\ell'}\widehat{q}_{\ell\ell'}
q_{\ell\ell}+\sum_{\ell}\widehat{m}_{\ell}m_{\ell}+F(\bsy{q},\bsy{m})+
S(\widehat{\bsy{q}},\widehat{\bsy{m}})]}
\end{equation}
where ($\alpha=M/N,\kappa\equiv\overline{(\beta_i^\mu-\alpha_i^\mu)^2}$)
\begin{eqnarray*}
F(\bsy{q},\bsy{m})=\alpha\log\int
\left[\prod_{\ell=1}^r\frac{dc_\ell d\widehat{c}_\ell}{2\pi}\right]
e^{i\sum_{\ell=1}^r \widehat{c}_\ell c_\ell +\ii g\sum_{\ell=1}^r
\widehat{c}_\ell
m_{\ell}-\frac{\kappa}{2}\sum_{\ell,\ell'}\widehat{c}_\ell
\widehat{c}_{\ell'} q_{\ell\ell'}}\\
\hspace{2cm}\times\prod_{\ell=1}^r\left[e^{-\beta}+(1-e^{-\beta})\Theta
(c_\ell)\right]\\ 
S(\widehat{\bsy{q}},\widehat{\bsy{m}})=\frac{1}{N}\log
\Tr_{\{\bsy{s}_\ell\}}\,e^{-\sum_{\ell\leq\ell'}
\widehat{q}_{\ell\ell'}\sum_{i=1}^N s_{i\ell}
s_{i\ell'}-\sum_{\ell}\widehat{m}_{\ell}\sum_{i=1}^N s_{i\ell}}
\end{eqnarray*}

In the limit $N\to\infty$, the integral (\ref{derp}) is evaluated by a
saddle-point method. Let us consider the saddle-point equation for the
order parameter $\widehat{q}_{\ell\ell'}$ ($\ell\neq\ell'$):
\begin{equation}\fl
\widehat{q}_{\ell\ell'}=\kappa\alpha\frac{\int
\prod_{\ell}\frac{dc_\ell d\widehat{c}_\ell}{2\pi}~ \widehat{c}_\ell
\widehat{c}_{\ell'}~e^{i\sum_{\ell=1}^r \widehat{c}_\ell c_\ell+\ii g\sum_{\ell}
\widehat{c}_\ell
m_{\ell}-\frac{\kappa}{2}\sum_{\ell,\ell'}\widehat{c}_\ell
\widehat{c}_{\ell'}
q_{\ell\ell'}}\prod_{\ell=1}^r\left[e^{-\beta}+(1-e^{-\beta})\Theta
(c_\ell)\right]}{\int \prod_{\ell}\frac{dc_\ell
d\widehat{c}_\ell}{2\pi}~e^{i\sum_{\ell=1}^r \widehat{c}_\ell
c_\ell+\ii g\sum_{\ell} \widehat{c}_\ell
m_{\ell}-\frac{\kappa}{2}\sum_{\ell,\ell'}\widehat{c}_\ell
\widehat{c}_{\ell'}
q_{\ell\ell'}}\prod_{\ell=1}^r\left[e^{-\beta}+(1-e^{-\beta})\Theta
(c_\ell)\right]}
\end{equation}
The argument of the exponential in the denominator can be linearized
by a Hubbard-Stratonovich transformation. Once this is done, it is
clear that the remaining integrals over $c_\ell$ and
$\widehat{c}_\ell$ factorize and one easily sees that in the replica
limit $r\to 0$ the denominator tends to $1$ and we can neglect it. It
then remains to evaluate the denominator with the replica-symmetric
(RS) Ansatz
\begin{eqnarray}
q_{\ell\ell'}=(Q-q)\delta_{\ell\ell'}+q~~~~~~~~~~~~ m_\ell=m\\
\widehat{q}_{\ell\ell'}=(\widehat{Q}-\widehat{q})\delta_{\ell\ell'}+\widehat{q}
~~~~~~~~~~~~ \widehat{m}_\ell=\widehat{m}
\end{eqnarray}
in the replica limit $r\to 0$. With standard manipulations one finds
\begin{equation}
\widehat{q}=-\kappa \alpha\int dx ~\mathcal{G}_x(0,\kappa
q)\left(\frac{(1-e^{-\beta})\frac{1}{\sqrt{2\pi \kappa(Q-q)}}
e^{-\frac{(gm+x)^2}{2\kappa(Q-q)}}}{e^{-\beta}+(1-e^{-\beta})
H\left(\frac{gm+x}{\sqrt{\kappa(Q-q)}}\right)} \right)^2
\end{equation}
where $\mathcal{G}_x(a,b)$ denotes a Gaussian distribution for the
variable $x$ with mean $a$ and variance $b$. while
$H(x)=\frac{1}{2}\erfc(x/\sqrt{2})$.

Along similar lines one can derive the corresponding equations for the
remaining order parameters:
\begin{eqnarray}
\widehat{Q}=-\frac{\kappa \alpha}{2}\int dx~ \mathcal{G}_x(0,\kappa
q)\frac{(1-e^{-\beta})\partial_x
\mathcal{G}_x(-gm,\kappa(Q-q))}{e^{-\beta}+(1-e^{-\beta})
H\left(\frac{gm+x}{\sqrt{\kappa(Q-q)}}\right)}\\ \widehat{m}=
g\alpha\int dx~ \mathcal{G}_x(0,\kappa
q)\frac{(1-e^{-\beta})\frac{1}{\sqrt{2\pi \kappa(Q-q)}}
e^{-\frac{(gm+x)^2}{2\kappa(Q-q)}}}{e^{-\beta}+(1-e^{-\beta})
H\left(\frac{gm+x}{\sqrt{\kappa(Q-q)}}\right)}\\ q=\int
dx\,\mathcal{G}_x(0,1)\left(\frac{\int
ds\,e^{-\left(\widehat{Q}-\frac{\widehat{q}}{2}\right) s^2+
(ix\sqrt{\widehat{q}} -\widehat{m}) s}s}{\int
ds\,e^{-\left(\widehat{Q}-\frac{\widehat{q}}{2}\right) s^2+
(ix\sqrt{\widehat{q}} -\widehat{m}) s}}\right)^2\\ Q=\int
dx\,\mathcal{G}_x(0,1)\left(\frac{\int
ds\,e^{-\left(\widehat{Q}-\frac{\widehat{q}}{2}\right) s^2+
(ix\sqrt{\widehat{q}} -\widehat{m}) s}s^2}{\int
ds\,e^{-\left(\widehat{Q}-\frac{\widehat{q}}{2}\right) s^2+
(ix\sqrt{\widehat{q}} -\widehat{m}) s}}\right)\\ m=\int
dx\,\mathcal{G}_x(0,1)\left(\frac{\int
ds\,e^{-\left(\widehat{Q}-\frac{\widehat{q}}{2}\right) s^2+
(ix\sqrt{\widehat{q}} -\widehat{m}) s}s}{\int
ds\,e^{-\left(\widehat{Q}-\frac{\widehat{q}}{2}\right) s^2+
(ix\sqrt{\widehat{q}} -\widehat{m}) s}}\right)
\end{eqnarray}

\section{The fully-connected limit of the cavity theory for the diluted model}

In this Appendix we study the large connectivity limit of the cavity
theory and derive equations for the relevant order parameters in this
limit, in order to show that these equations are equivalent to those
obtained in the revisited replica theory presented above. The first
problem is to study the cavity distributions (\ref{zoppola}) and
(\ref{zoppolo}) in the fully connected limit for a fixed disorder
realization. Next, we shall perform the disorder average.

Let us begin from (\ref{zoppolo}), which we re-cast as
\begin{eqnarray}
q_{(i)}^\mu(c)&=&
\Tr_{\bsy{s}_{j\in\mu\backslash i}}
\l[\prod_{j\in\mu\backslash i} p_j^{(\mu)}(s_j)\r]
\delta\l[c-\frac{1}{\sqrt{K}}\sum_{j\in\mu\backslash i} s_j
\xi_j^\mu(\rho)\r]\\\label{portio}
&=&\int\frac{d\widehat{c}}{2\pi}e^{\ii c \widehat{c}}
\prod_{j\in\mu\backslash i}
\Tr_{s_j} p^{(\mu)}_j(s_j) e^{-\frac{\ii}{\sqrt{K}}\widehat{c}
s_j\xi_j^\mu(\rho)}
\end{eqnarray}
where we introduced a re-scaling factor $1/\sqrt{K}$ for the fields
($K$ being the average connectivity). For $K\gg 1$ we can expand the
exponential. Keeping terms up to the second order we find
\begin{equation}\label{martio}\fl
\prod_{j\in\mu\backslash i}
\Tr_{s_j} p^{(\mu)}_j(s_j) e^{-\frac{\ii}{\sqrt{K}}\widehat{c}
s_j\xi_j^\mu(\rho)}=
\exp\left[-\frac{\ii}{\sqrt{K}}\widehat{c} 
\sum_{j\in \mu\backslash i}\xi_j^\mu(\rho) m_j^{(\mu)}- 
\frac{\widehat{c}^2}{2 K} 
\sum_{j\in \mu\backslash i}
[\xi_j^\mu(\rho)]^2 \Delta_j^{(\mu)}\r]
\end{equation}
where
\begin{eqnarray}
m_i^{(\mu)}=\avg{s_i}_{(\mu)}\\
\Delta_i^{(\mu)}=\avg{s_i^2}_{(\mu)}-\avg{s_i}_{(\mu)}^2\\
\avg{f(s_i)}_{(\mu)}=\Tr_s p_i^{(\mu)}(s)f(s_i)
\end{eqnarray}
Inserting (\ref{martio}) into (\ref{portio}) and integrating over
$\widehat{c}$ one sees that for $K\gg 1$
\begin{equation}\label{gauaa}
q_{(i)}^\mu(c)=\mathcal{G}_c\l[\frac{1}{\sqrt{K}}
\sum_{j\in \mu\backslash i}\xi_j^\mu(\rho) m_j^{(\mu)},
\frac{1}{K}\sum_{j\in \mu\backslash i}[\xi_j^\mu(\rho)]^2 
\Delta_j^{(\mu)}\r]
\end{equation}
where as before $\mathcal{G}_x(a,b)$ denotes a Gaussian distribution
for the variable $x$ with mean $a$ and variance $b$.

The limit of (\ref{zoppola}) is a bit more complicated. For a start,
we re-write it by introducing the usual re-scaling:
\begin{eqnarray}
p_i^{(\mu)}(s)&=&\frac{1}{Z_s}\Tr_{\bsy{c}^{\nu\in i\backslash \mu}}
\l[\prod_{\nu\in i\backslash \mu}q_{(i)}^\nu(c^\nu)\r]
e^{-\beta\sum_{\nu\in
i\backslash \mu}\Theta\l[-c_{(i)}^\nu-\frac{1}{\sqrt{K}}s \xi_i^\nu(\rho)\r]}\\
&=&\frac{1}{Z_s} \prod_{\nu\in i\backslash \mu}
\Tr_c q_{(i)}^\nu(c) 
e^{-\beta 
\Theta\l[-c-\frac{1}{\sqrt{K}}s \xi_i^\nu(\rho)\r]}\label{zuo}
\end{eqnarray}
We focus on
\begin{equation}
\fl \Tr_c q_{(i)}^\nu(c) 
e^{-\beta 
\Theta\l[-c-\frac{1}{\sqrt{K}}s \xi_i^\nu(\rho)\r]}=
\Tr_c q_{(i)}^\nu(c) \l[e^{-\beta}
+\l(1-e^{-\beta}\r)\Theta\l(c+\frac{1}{\sqrt{K}}s \xi_i^\nu(\rho)\r)\r]
\end{equation}
Now for any $K$ the Heaviside function can be disposed of via
\begin{eqnarray}
\Theta\left(c+\frac{1}{\sqrt{K}}s 
\xi^\nu_i(\rho)\right)=\Theta\left(c\right)+
\varepsilon^\nu(c,s)\\
\varepsilon^\nu(c,s)= \sum_{n\geq 1} 
\frac{1}{n!}\left(\frac{1}{\sqrt{K}}s\, 
\xi_i^\nu(\rho)\right)^{n}\delta^{(n-1)}\left(c\right)
\label{mustocco}
\end{eqnarray}
so that
\begin{eqnarray*}
 \Tr_c q_{(i)}^\nu(c) 
e^{-\beta 
\Theta\l[-c-\frac{1}{\sqrt{K}}s \xi_i^\nu(\rho)\r]}
&=&\Tr_c q_{(i)}^\nu(c) \l[e^{-\beta}+\l(1-e^{-\beta}\r)\Theta(c)\r]
\\
&+&\l(1-e^{-\beta}\r)\Tr_c q_{(i)}^\nu(c)\varepsilon^\nu(c,s)
\nonumber
\end{eqnarray*}
Factoring out a term $\Tr_c q_{(i)}^\nu(c)
\l[e^{-\beta}+\l(1-e^{-\beta}\r)\Theta(c)\r]$ one sees that
(\ref{zuo}) becomes
\begin{equation}
p_i^{(\mu)}(s)=\frac{1}{\widetilde{Z}_s}\prod_{\nu\in i\backslash \mu}
\l[1+\l(1-e^{-\beta}\r)\Tr_c \mathcal{Q}_{(i)}^\nu(c)
\varepsilon^\nu(c,s)\r]
\end{equation}
where $\widetilde{Z}_s$ is a new normalization factor and
\begin{equation}\label{tren}
\mathcal{Q}_{(i)}^\nu(c)=\frac{q_{(i)}^\nu(c)}{\Tr_c q_{(i)}^\nu(c)
\l[e^{-\beta}+\l(1-e^{-\beta}\r)\Theta(c)\r]}
\end{equation}
With the expression given in (\ref{mustocco}), we have
\begin{equation}
\Tr_c \mathcal{Q}_{(i)}^\nu(c)
\varepsilon^\nu(c,s)=\sum_{n\geq1} \frac{1}{n!}\left(\frac{1}{\sqrt{K}}s\, 
\xi_i^\nu(\rho)\right)^{n} (-1)^{n-1}\mathcal{Q}_{(i)}^{\nu,(n-1)}(0)
\end{equation}
For $K\gg 1$ we can truncate the above expansion after the first two
terms and approximate the resulting expression with an exponential.
This gives us
\begin{equation}\label{gaua}
p_i^{(\mu)}(s)=\frac{e^{A^{(\mu)}_i s-B^{(\mu)}_i s^2}}{
\Tr_s e^{A^{(\mu)}_i s-B^{(\mu)}_i s^2}}
\end{equation}
with
\begin{eqnarray}
A^{(\mu)}_i= \left(1-e^{-\beta}\right)\frac{1}{\sqrt{K}}\,
\sum_{\nu\in i\backslash \mu} \xi_i^\nu(\rho)\mathcal{Q}_{(i)}^\nu(0)\\
B^{(\mu)}_i= \left(1-e^{-\beta}\right) \frac{1}{2K}\sum_{\nu\in i\backslash \mu} 
\l[\xi_i^\nu(\rho)\r]^2\left( \mathcal{Q}^{\nu,(1)}_{(i)}(0)+
\left(1-e^{-\beta}\right)[\mathcal{Q}_{(i)}^\nu(0)]^2\right)
\end{eqnarray}

(\ref{gauaa}) and (\ref{gaua}) represent the $K\gg 1$ limits of the
cavity distributions for a fixed disorder sample. 

Recalling (\ref{defa}) and (\ref{defa2}), we can now evaluate averages
over the quenched disorder. To this aim, we set $\rho=1+g/\sqrt{K}$ so
that
\begin{equation}
\xi_i^\mu(\rho)=\beta_i^\mu-\alpha_i^\mu-\frac{g}{\sqrt{K}}(1+\alpha_i^\mu)
\end{equation}

Computing the statistics of the quantities $A^{(\mu)}_i$ and
$B^{(\mu)}_i$ over disorder one obtains (with $\alpha=K/N$)
\begin{eqnarray}
\overline{A^{(\mu)}_i}\equiv A=-g\alpha \left(1-e^{-\beta}\right)
F\\\ovl{(A^{(\mu)}_i-A)^2}\equiv\sigma_{A}=\alpha\kappa
\left(1-e^{-\beta}\right)^2 D\\ \overline{B^{(\mu)}_i}\equiv B=\alpha
\kappa\left(1-e^{-\beta}\right)\left[C+\left(1-e^{-\beta}\right)D\right]
\end{eqnarray}
($B^{(\mu)}_i$ is sample-independent in the fully connected limit
$K\to\infty$, so we neglect its fluctuations) where we used the
following shorthands:
\begin{eqnarray}
C=\overline{\mathcal{Q}^{\nu,(1)}_{(i)}(0)}\\
D=\overline{[\mathcal{Q}^{\nu}_{(i)}(0)]^2}\\
F=\overline{\mathcal{Q}^{\nu}_{(i)}(0)}\\
\kappa\equiv\overline{(\beta_i^\mu-\alpha_i^\mu)^2}
\end{eqnarray}
This implies that
\begin{equation}
\overline{[\avg{f(s_i)}_{(\mu)}]^k}=\int
dH~\mathcal{G}_{H}(A,\sigma_A)
\left[\frac{\Tr_s f(s) e^{Hs-B s^2}}{\Tr_s 
e^{ H s-B s^2}}\right]^k
\end{equation}
We now must evaluate $C$, $D$ and $F$, recalling
$\mathcal{Q}^{\nu}_{(i)}(c)$ as defined in Equations (\ref{tren}) and
(\ref{gauaa}). The statistics of $h=\frac{1}{\sqrt{K}} \sum_{j\in
\mu\backslash i}\xi_j^\mu(\rho) m_j^{(\mu)}$ is given by
\begin{eqnarray}
\ovl{h}=-gM,~~~~~~~~~~~~~~~~~~~M=\ovl{m_i^{(\mu)}}\\
\ovl{h^2}=g^2 M^2+\kappa q,~~~~~~~~~~~q=\ovl{\avg{s_i}_{(\mu)}^2}\\
\ovl{(h-\ovl{h})^2}=\kappa q
\end{eqnarray}
For the variance of $\mathcal{Q}^{\nu}_{(i)}(c)$ we obtain instead
\begin{equation}
\ovl{\frac{1}{K}\sum_{j\in \mu\backslash i}[\xi_j^\mu(\rho)]^2 
\Delta_j^{(\mu)}}=\kappa(Q-q),~~~~~~~
Q=\ovl{\avg{s_i^2}_{(\mu)}}
\end{equation}
We can therefore write
\begin{eqnarray}\fl
F=\int dh\,\mathcal{G}_h(-g M,\kappa
q)\frac{\mathcal{G}_{c=0}\left[h,\kappa(Q-q)\right]}{\int
dc\,\mathcal{G}_{c}\left[h,\kappa(Q-q)\right]\left[
e^{-\beta}+\left(1-e^{-\beta}\right)\Theta\left(c\right)\right]}\\
\fl
D=\int dh\,\mathcal{G}_h(-g M,\kappa
q)\left[\frac{\mathcal{G}_{c=0}\left[h,\kappa(Q-q)\right]}{\int
dc\,\mathcal{G}_{c}\left[h,\kappa(Q-q)\right]\left[
e^{-\beta}+\left(1-e^{-\beta}\right)\Theta\left(c\right)\right]}\right]^2\\
\fl
C=\int dh\,\mathcal{G}_h(-g M,\kappa
q)\frac{\mathcal{G}^{(1)}_{c=0}\left[h,\kappa(Q-q)\right]}{\int
dc\,\mathcal{G}_{c}\left[h,\kappa(Q-q)\right]\left[
e^{-\beta}+\left(1-e^{-\beta}\right)\Theta\left(c\right)\right]}
\end{eqnarray}
These form a closed system together with the equations for $M$, $Q$
and $q$:
\begin{eqnarray}
M=\overline{\avg{s_i}_{(\mu)}}=\int
dH~\mathcal{G}_{H}(A,\sigma_A)
\left[\frac{\Tr_s s~ e^{Hs-B s^2}}{\Tr_s 
e^{ H s-B s^2}}\right]\\
Q=\overline{\avg{s_i^2}_{(\mu)}}=\int
dH~\mathcal{G}_{H}(A,\sigma_A)
\left[\frac{\Tr_s s^2~ e^{Hs-B s^2}}{\Tr_s 
e^{ H s-B s^2}}\right]\\
q=\overline{\avg{s_i}_{(\mu)}^2}=\int
dH~\mathcal{G}_{H}(A,\sigma_A)
\left[\frac{\Tr_s s~ e^{Hs-B s^2}}{\Tr_s 
e^{ H s-B s^2}}\right]^2
\end{eqnarray}
With minor manipulations, these equations can be precisely identified
with the equations obtained by the replica approach in RS
approximation in the previous section.

\end{document}